\documentclass[fleqn,usenatbib]{mnras}

\usepackage{newtxtext,newtxmath}
\usepackage[T1]{fontenc}
\usepackage{xcolor}
\usepackage{pdfpages}
\usepackage{fix-cm}
\usepackage{float}
\usepackage{svg}

\svgsetup{inkscapelatex=false}
\usepackage{graphicx}

\usepackage{ragged2e}
\usepackage[utf8]{inputenc}
\usepackage{booktabs}

\usepackage[T1]{fontenc}
\usepackage{multirow}

\DeclareRobustCommand{\VAN}[3]{#2}
\let\VANthebibliography\thebibliography
\def\thebibliography{\DeclareRobustCommand{\VAN}[3]{##3}\VANthebibliography}

\usepackage{graphicx}	
\usepackage{amsmath}	
\usepackage{float}

\usepackage{url}
\usepackage{xcolor}
\usepackage{lineno}

\title{Content-Based Image Retrieval Using COSFIRE Descriptors with application to Radio Astronomy} 

\author[S.Ndung'u et al.]{
Steven Ndung'u,$^{1,2}$\thanks{E-mail: 26846578@sun.ac.za}
Trienko Grobler,$^{1}$
Stefan J. Wijnholds$^{1,3}$
and George Azzopardi$^{2}$
\\
$^{1}$University of Stellenbosch, Cnr Banhoek Road \& Joubert Street, Stellenbosch, 7600, South Africa\\
$^{2}$University of Groningen, Nijenborgh 9, 9712 CP, Groningen, The Netherlands\\
$^{3}$ASTRON, Oude Hoogeveensedijk 4, 7991 PD. Dwingeloo, The Netherlands
}

\date{Accepted xxx. Received xxx; in original form xxx}

\pubyear{2024}
\begin{document}
\label{firstpage}
\pagerange{\pageref{firstpage}--\pageref{lastpage}}
\maketitle

\begin{abstract}
The morphologies of astronomical sources are highly complex, making it essential not only to classify the identified sources into their predefined categories but also to determine the sources that are most similar to a given query source. Image-based retrieval is essential, as it allows an astronomer with a source under study to ask a computer to sift through the large archived database of sources to find the most similar ones. This is of particular interest if the source under study does not fall into a “known” category (anomalous). Our work uses the trainable COSFIRE (Combination of Shifted Filter Responses) approach for image retrieval. COSFIRE filters are automatically configured to extract the hyperlocal geometric arrangements that uniquely describe the morphological characteristics of patterns of interest in a given image; in this case astronomical sources. This is achieved by automatically examining the shape properties of a given prototype source in an image, which ultimately determines the selectivity of a COSFIRE filter. We further utilize hashing techniques, which are efficient in terms of required computation and storage, enabling scalability in handling large data sets in the image retrieval process. We evaluated the effectiveness of our approach by conducting experiments on a benchmark data set of radio galaxies, containing 1,180 training images and 404 test images. Notably, our approach achieved a mean average precision of 91\% for image retrieval, surpassing the performance of the competing DenseNet-based method. Moreover, the COSFIRE filters are significantly more computationally efficient, requiring $\sim\!14\times$ fewer operations than the DenseNet-based method.

\end{abstract}



\section{Introduction}

Image retrieval is the process by which a computer system is designed for indexing, searching, browsing, and retrieving images from large databases. Astronomy is characterized by large volumes of data, exemplified by the Sloan Digital Sky Survey (SDSS)\footnote{https://classic.sdss.org/home.php}, which catalogs over 500 million astronomical objects to date \citep{abdurro2022seventeenth,almeida2023eighteenth}. Similarly, the LOw-Frequency ARray (LOFAR) survey has documented  $\sim$4.4 million radio sources \citep{shimwell2022lofar}, while the Square Kilometre Array (SKA) is anticipated to identify over 500 million additional radio sources \citep{norris2014ska}.  Such a system enables astronomers to input an image of a celestial object and quickly find similar sources in large, archived celestial source databases. This is of particular interest if the source of study does not fall into a "known" classification category, such as the double radio relic and odd radio circles that were discovered recently \citep{norris2022meerkat,koribalski2023meerkat}. Furthermore, in the efforts to disseminate scientific knowledge at all levels within the astronomy community by enhancing data archive accessibility \citep{tallada2020cosmohub} and through public citizen science projects, such as the Radio Galaxy Zoo \citep{banfield2015radio}, a simplified and intuitive query system is essential. Such a system eliminates the need for users to possess in-depth knowledge of underlying astronomical morphological characteristics or naming conventions. This approach aims to provide democratized access to astronomical data for the astronomy community in an efficient and scalable manner.

Content-based image retrieval (CBIR) is a framework used to search an image archive or catalogue to retrieve images with matching visual content to a specified query image. CBIR approaches rely on extracting global morphological features (such as colour, size, texture, shape, and spatial positions) and low-level features (such as blobs, corners and edges). When a query image is introduced, the system computes the similarity between the stored features and those of the query image, extracted using the same methods, to determine the most similar matches \citep{Csillaghy_2000,Long2003,AbdElAziz2017AutomaticDO}. While CBIR features provide useful information, they are limited in their capacity to capture the semantic content of images; images with similar low-level features may differ significantly in terms of their semantics \citep{DBLP:conf/icisp/AbiouiIBM18}. Furthermore, in the inference phase, since the images are represented as real-valued features, there are high computational and storage costs associated with them, particularly when managing large image databases. This combination of the semantic gap and the high computation/storage costs poses significant challenges for traditional CBIR approaches, especially when working with large-scale image databases \citep{chao2022deep}. To address these challenges we propose a lightweight approach to image retrieval that involves COSFIRE filter descriptors and a multi-layer perceptron (MLP). An approach that is not only efficient and rotation-tolerant but also robust against small and imbalanced training sets. 

In this paper we illustrate the usefulness of our technique by applying it to radio sources. This process would help radio astronomers advance the understanding of the formation of different radio galaxies, the evolution of galaxies, the accretion processes, and the interactions of their jets with their surrounding environment. Our approach encompasses a three-phase image retrieval pipeline.  Phase one involves feature extraction using trainable COSFIRE filters \citep{azzopardi2012trainable, 10.1093/mnras/stae821}. The COSFIRE filters approach is inspired by the neurobiological mechanisms observed in the human visual system and is analogous to population coding, the mechanism by which neurons in the mammalian brain encode visual information \citep{pasupathy1999responses,pasupathy2002population}. It has been successfully applied to various computer vision tasks, including object recognition \citep{azzopardi2016gender} and image classification \citep{GECER2017165,10.1093/mnras/stae821}. The configuration of COSFIRE filters is an automated process designed to capture and extract hyperlocal patterns that form the image signature that is characteristic of radio source morphology. This process is achieved through the examination of geometric characteristics of a selected prototype pattern within the image, which ultimately defines a COSFIRE filter's selectivity. 

In phase two, we apply a MLP for image hashing, utilizing the features generated by the COSFIRE filter responses. The hashing methodology employs an artificial neural network architecture to concurrently learn feature representations and hash functions, thus enhancing the accuracy and separability of the radio galaxies. During the learning process, we use a discriminative loss function on image pairs to cluster similar images and separate dissimilar ones, creating a feature representation space that effectively approximates the semantic relationships of the images.

\begin{figure}
 \centering
 \footnotesize
\includegraphics[trim=0.7cm 0.6cm 0.9cm 0.6cm,clip,width=\columnwidth]{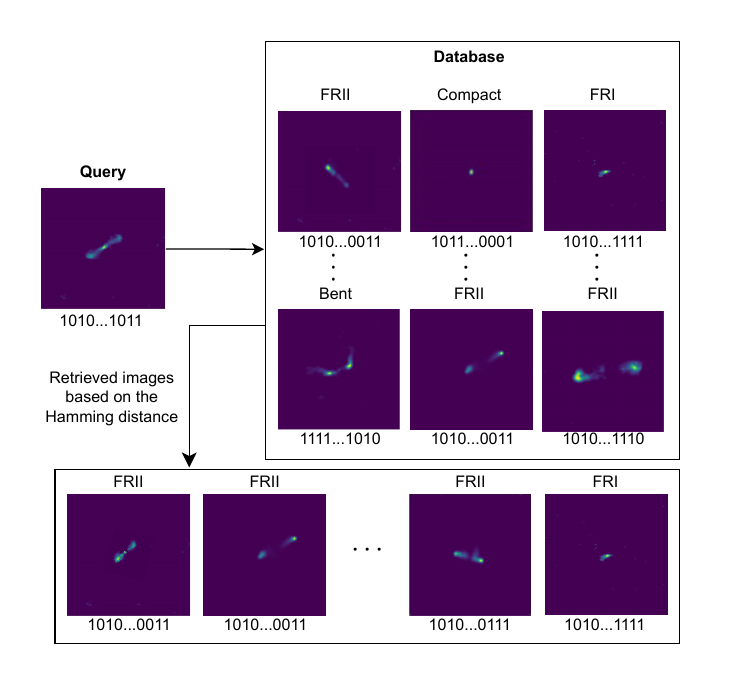}
\caption{The image retrieval process. This framework illustrates the image retrieval process for a given query image, assuming all images have been binarized using a hashing function. It involves computing the Hamming distance to measure the similarity between the query image and each image in the reference set. The outcome is a selection of the top N images that most closely match the query. In this case, we use four classes of radio galaxies, classified based on their relative source positions and their low- and high-luminosity regions: FRI (center-brightened sources and sometimes diffuse), FRII (sources with two bright lobes, spatially separated from the core, which outshine both the core and the jets), Bent (sources with jets that are inclined at an angle), and Compact (point-like sources).
}
\label{img:intro_image}
\end{figure}

Finally, phase three involves binarization and image retrieval. In this step, the binary-like outputs from the MLP hash layer are converted into compact binary codes. The similarity between these binary representations is then measured using the Hamming distance, enabling efficient retrieval of the most similar images from large-scale databases. 
The image retrieval process is illustrated using a schematic overview in Figure \ref{img:intro_image}. Furthermore, we perform comparative analyses between our study and a closely related work that utilized DenseNet-161 \citep{10265687}, employing the same data set and loss function.

The subsequent sections of the paper are organized in the following manner: Section \ref{sec: relatedworks} examines the related state-of-the-art techniques in image retrieval. Section \ref{sec: data} explains the data set utilised in our study. Section \ref{sec: methods} presents the process of obtaining COSFIRE descriptors, the MLP architecture, the loss function, the binarization process, and the image retrieval procedure. Section \ref{sec: evaluation} outlines the evaluative measures used to determine the performance of our technique. Section \ref{sec: experiments_and_results} presents the experiments and the results. The discussion of the findings in comparison to the existing literature is presented in Section ~\ref{sec: discussion}. Finally, the conclusion is provided in Section ~\ref{sec: conclusion}.

\section{Related works}
\label{sec: relatedworks}

CBIR methods have been increasingly applied to astronomical data, addressing the challenges of handling large, complex data sets. Early work on CBIR in astronomy includes \citet{1998ASPC..145..429P}'s Query-By-Example (QUBE) system, which pioneered the use of artificial neural networks for image matching in astronomical archives. Additionally, \citet{ardizzone1996suitability} introduced a technique using Normalized Axial Moments for shape description, coupled with multi-resolution compression to reduce computational load. While that approach offered rotation and scaling invariance - crucial for astronomical sources - its effectiveness was heavily dependent on compression quality, risking information loss if not carefully implemented. \citet{Csillaghy_2000} proposed a two-phase indexing approach for solar radio spectrograms, combining texture summarization and feature generation using self-organizing maps. While innovative in handling diffuse and noisy data, this method faced computational challenges and struggles with imbalanced class distributions. Advancing the field, \citet{abd2017automatic} developed a comprehensive CBIR method for galaxy-type detection, integrating shape, color, and texture features with a Binary Sine Cosine algorithm for feature selection. Their approach, tested on the EFIGI (``Extraction de Formes Idéalisées de Galaxies en Imagerie'') catalogue \citep{baillard2011efigi}, demonstrated superior performance in precision, recall, and efficiency compared to particle swarm optimization and genetic algorithms. These diverse approaches highlight the evolution of CBIR techniques in astronomy, from early attempts at intuitive querying to sophisticated methods combining feature extraction, selection, and retrieval, addressing the unique challenges posed by astronomical data.

Recently, several studies have demonstrated the effectiveness of machine and deep learning in various image classification \citep{Lukic_2019,2019MNRAS.488.3358T,samudre2022data,ndung2023advances} and retrieval tasks \citep{liu2016deep,walmsley2022practical, 10265687}. \citet{variawa2022transfer}  utilized deep metric learning, leveraging the ResNet50 architecture \citep{he2016deep} combined with transfer learning, for the classification and retrieval of galaxy morphologies. This approach was applied to data sets that were crowd-sourced, namely Galaxy Zoo 2 \citep{willett2013galaxy}, as well as those labeled by experts, EFIGI \citep{de2013third}. Their findings demonstrated that metric learning is effective, achieving state-of-the-art outcomes in both classification and image retrieval tasks, even with the challenges posed by limited and imbalanced expert-labeled data sets. \citet{walmsley2022practical} pretrained a deep learning network on astronomy-specific data, the Dark Energy Camera Legacy Survey DR5 \citep{dey2019overview}, as opposed to using terrestrial ImageNet images \citep{deng2009imagenet} to learn general galaxy representations for downstream tasks such as classification, image retrieval, and anomaly detection. They found that the image representations learned could generalize better compared to ImageNet pretraining for galaxy-specific tasks. However, it was also noted that the performance depended on the quality and diversity of the initial training data used. Moreover, deep hashing based on convolutional neural networks can be trained to generate hash representations that preserve the semantic similarity between images \citep{liu2016deep,chao2022deep}. These binary representations enable rapid nearest neighbor searches, making deep hashing particularly suitable for applications involving extensive image databases, compared to traditional feature-based CBIR approaches. In radio astronomy, \citet{10265687} demonstrated that a transfer learning-based deep CNN (DenseNet-161) model can effectively extract distinctive features, enabling efficient and accurate retrieval of radio galaxies.

From the review of previous work on image retrieval in astronomy, there is a clear need for methods that are robust enough to capture the diverse and complex morphological characteristics of radio galaxies, including rotational invariance, while also being efficient in terms of computational and storage resources. To our knowledge, our previous work \citep{10265687} and this study are the first to apply machine and deep learning approaches to radio data in radio astronomy. In this study, we investigate the COSFIRE approach, which has demonstrated robustness and computational efficiency in classification on the data set used for this image retrieval task \citep{10.1093/mnras/stae821}.

\section{Data}

\label{sec: data}

The data set used in this paper is composed of $\sim$2,000 samples with four distinct classes of radio galaxies, namely Compact (406 samples), FRI (389 samples), FRII (679 samples) and  Bent (508 samples), see Table~\ref{tab:original_dataset_distrb}. It was compiled and processed by \citet{samudre2022data}. They carefully selected resolved radio galaxies from various catologues for each radio galaxy class.  For instance, Compact radio galaxies were obtained from the Combined NVSS–FIRST\footnote{National Radio Astronomy Observatory  Very Large Array  Sky Survey -  Faint Images of the Radio Sky} galaxies catalogue (CoNFIG) \citep{gendre2008combined,10.1111/j.1365-2966.2010.16413.x} and  FR0CAT catalogue \citep{baldi2018fr0cat}; FRI radio galaxies were obtained from the CoNFIG and FRICAT catalogues \citep{capetti2017fricat};  FRII radio galaxies were obtained from the CoNFIG and FRIICAT catalogues, and finally Bent radio galaxies from the Proctor catalogue \citep{proctor2011morphological}.

The single pre-processing operation of the images involved the application of sigma-clipping, setting a threshold of 3$\sigma$ \citep{aniyan2017classifying}. This method iteratively identifies and masks pixels that deviate significantly from local background levels, effectively reducing noise while preserving real radio source pixels. The process was carefully tuned (to 3$\sigma$ threshold) to ensure that genuine astronomical sources were not inadvertently removed.

\begin{table}
\footnotesize
\centering
\caption{\textbf{The distribution of the \textit{original} data set across the training, validation, and test sets, along with their respective sizes.}}
\label{tab:original_dataset_distrb}
\begin{tabular}{@{}p{2.4cm}p{0.7cm}p{0.7cm}p{0.9cm}p{1cm}p{0.7cm}@{}}
\toprule
\textbf{Source catalog} & \textbf{Type}  & \textbf{Train} & \textbf{Valid}  & \textbf{Test} & \textbf{Total} \\
\midrule 
Proctor & Bent  & 305  & 100 & 103 & 508 \\
\addlinespace
FR0CAT \& CoNFIG & Compact  & 226 & 80   & 100 & 406\\
\addlinespace
FRICAT \& CoNFIG & FRI  & 215 & 74 & 100  & 389 \\
\addlinespace
FRIICAT \& CoNFIG& FRII  & 434  & 144 & 101 & 679\\
\midrule 
Total & - & 1180  & 398 &  404 & 1982 \\
\bottomrule
\end{tabular}
\end{table}

\section{Methods}
\label{sec: methods}

This section outlines our methodology for image retrieval using COSFIRE filter descriptors. We detail the process of obtaining the COSFIRE filter descriptors, the architecture of the hashing model, and the loss function employed to achieve effective image retrieval.

\subsection{COSFIRE descriptors}

We provide an overview of the trainable COSFIRE filter technique and the method for deriving radio galaxy descriptors for image retrieval. For comprehensive details on COSFIRE filters, we refer the reader to \citet{10.1093/mnras/stae821} and \citet{azzopardi2012trainable}.

\begin{figure*}
 \centering
 \footnotesize
\includegraphics[trim={0 0.50cm 0 0.50cm},clip,width=\textwidth]{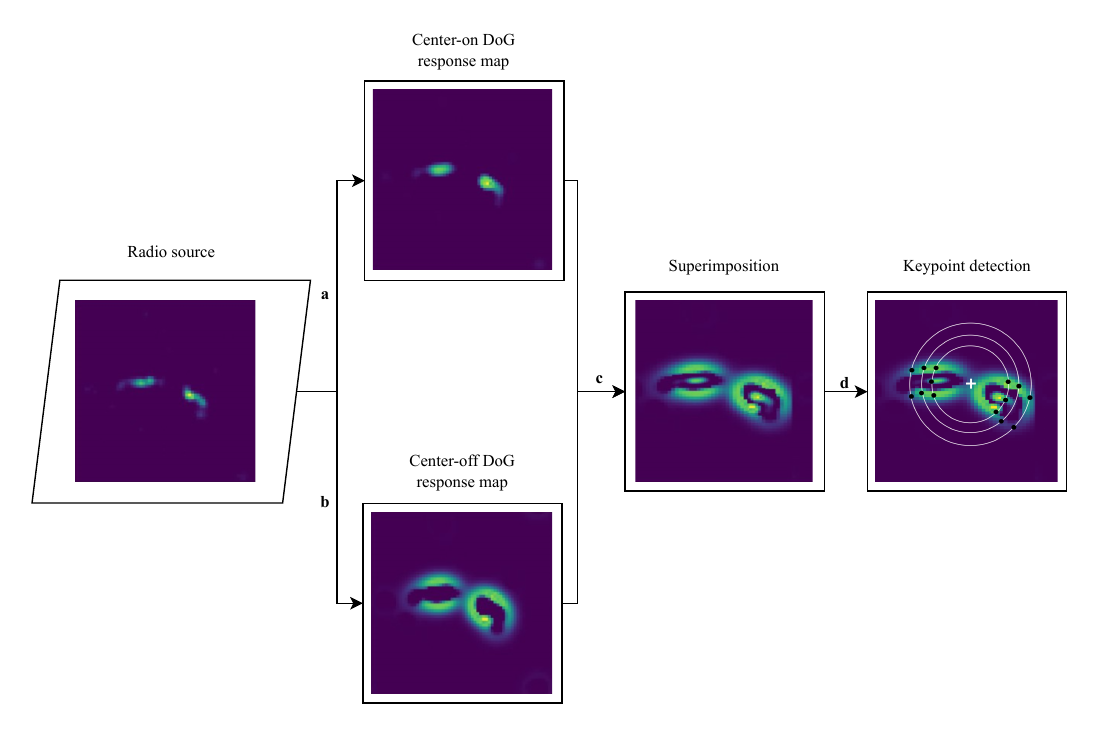}
\caption{Configuration of a COSFIRE filter. This figure shows four key steps indicated by the letters a-d: a) To the given $150\times150$ pixel image (Bent class in this case), perform center-on DoG convolution, and b) center-off DoG convolution; c) Superimpose the center-on and center-off DoG response maps and d) Then consider $k$ (3 for this illustration) concentric circles and identify the locations that show local maximum DoG responses along the these circles with respect to the center - 13 points are identified in this example. The DoG response maps are obtained with DoG functions with  a standard deviation of three ($\sigma$=3) for the outer Gaussian function. The  “+'' marker indicates the centre of the image.}
\label{fig: configuration}
\end{figure*}

The COSFIRE filter approach is adept at forming descriptors for complex patterns, such as radio galaxies. This method is grounded in the analysis of shape properties and is configured automatically by applying a bank of center-on and center-off Difference-of-Gaussians (DoG) filters to a given image. A DoG function is an approximation of the second-order derivative of a Gaussian function (Laplacian of Gaussian) with the advantage of being separable, thus making convolutions much more efficient. A center-on DoG filter responds to bright-to-dark contrast changes, while a center-off DoG filter responds to dark-to-bright contrast changes. These filters, which differ in polarity (center-on or center-off) and scale, are convolved with the input image to produce DoG response maps.

The configuration of a COSFIRE filter is accomplished through a keypoint detection process that entails the consideration of multiple concentric circles centered on the pattern of interest and identifying the strongest local DoG filter responses along these circles. The number of circles and their radii are hyperparameters. For each salient point $i$ detected, its distance $\rho_i$ and angular position $\phi_i$ relative to the center of the prototype pattern of interest are recorded, along with the polarity $\delta_i$ and the standard deviation $\sigma_i$ of the outer Gaussian function of the DoG filter that elicited the strongest response at that location\footnote{The standard deviation of the inner Gaussian function is half that of the outer Gaussian function}. This information is encoded into a list of tuples that characterize the structure of a COSFIRE filter $C_f$:

\begin{equation}
    C_f = \{(\rho_i,\phi_i,\sigma_i,\delta_i)~|~i=1\dots n\}
\end{equation}

\noindent where $n$ denotes the number of detected keypoints along the considered concentric circles.

For our data set, the galaxies—our patterns of interest—are already centered in the images. Therefore, we use the image center as the reference point for configuring the COSFIRE filters. The configuration process of the keypoint detection using the COSFIRE filter is illustrated in Fig.~\ref{fig: configuration}.

Once configured, the COSFIRE filter's response is determined in three steps. First the DoG response maps are computed for all unique pairs of $\delta$ and $\sigma$. Then, to introduce a degree of tolerance for the DoG responses' preferred positions, each response map is blurred with a Gaussian function. For each tuple $i$ in $C_f$ we take the respective blurred DoG response map and shift it in the direction opposite to the one defined by the polar coordinates $(\rho_i,\phi_i)$. In this way, all points of interests are aligned at the same location. The COSFIRE filter response is ultimately calculated using the geometric mean function, in each column along the stack of blurred and shifted DoG response maps. 

To ensure robustness against orientation changes, several COSFIRE filters are configured based on rotated versions of a single COSFIRE filter. This rotational invariance allows the filters to detect radio sources regardless of their orientation within the image. 

Multiple COSFIRE filters can be configured on patterns of interests drawn from the training set. Eventually, a COSFIRE descriptor is generated by applying all rotation-tolerant COSFIRE filters to a given image and extracting the maximum response from each COSFIRE filter. This yields a descriptor vector that encapsulates the image, effectively capturing the essential structure and patterns of the radio galaxy source. These descriptors were used in the classification of radio galaxies using the Support Vector Machine algorithm, yielding state-of-the-art results on the data set detailed in Section \ref{sec: data}. The detailed results can be found in the paper \citep{10.1093/mnras/stae821}.

\subsection{COSFIRE descriptors selection}
\label{sec:COSFIRE_descriptors_selection}

This work introduces a compact image hash code learning framework based on COSFIRE filter descriptors, designed for efficient similarity search and retrieval. We specifically leverage the training, validation, and test descriptors derived from the 26 hyperparameter sets\footnote{A hyperparameter set consists of five parameters: three used in the configuration (the standard deviation of the outer Gaussian function, $\sigma$; the set of radii $P$ for configuring COSFIRE filters; and the threshold $t_1$, which sets to zero all DoG responses below it) and two used in the application ($\sigma'$ and $\alpha$, which determine the standard deviation of the blurring function applied to the DoG response maps).} (COSFIRE configurations) identified in the classification methodology outlined in \citep{10.1093/mnras/stae821}. These descriptors were found to yield classification accuracies comparable to the best-performing hyperparameter set, with no statistically significant difference from the highest validation accuracy, as determined by a right-tailed Student's \textit{t}-test. This test was used to assess whether there was a significant increase in mean performance compared to a specified benchmark. The maximum average accuracy across these 26 COSFIRE configurations was achieved using a total of 372 COSFIRE filters on the validation set, distributed as 93 filters per class across the four radio source classes. Consequently, we employ the descriptors from these 93 COSFIRE filters per class as the foundation for our image hashing experiments, utilizing each of the 26 descriptor sets in our work.

As a single preprocessing step, L2 normalization is applied to the COSFIRE descriptors. L2 normalization involves scaling a vector to unit length by dividing it by its magnitude, which is the square root of the sum of its squared elements. This technique enhances the robustness of the descriptors against changes in illumination and contrast, ensuring consistent feature vector magnitudes. Consequently, it facilitates more accurate comparisons between feature descriptors from images under different conditions, thus boosting the effectiveness of downstream image recognition and retrieval tasks.

\subsection{The hashing model architecture}
\label{hashing_architecture}
A MLP hashing network is developed to learn binary hash representations that capture the semantic similarity of the images. This enables efficient matching of binary hash representations of database images for fast retrieval from the COSFIRE descriptor feature vectors. MLP is a type of artificial neural network consisting of fully connected layers. The neurons in the MLP typically use nonlinear activation functions, allowing the network to learn complex patterns in the COSFIRE descriptors. We train a MLP network architecture using the loss function described in Section~\ref{scn: loss_function}, which guides the neural network to map the input descriptors into a latent feature space that enhances the similarities and differences crucial for effective image class separation. The resulting model provides a pipeline from raw COSFIRE image descriptors to a $k$-bit compact hash code suitable for efficient image retrieval and matching. The architecture is constructed as a sequential stack comprising linear layers, batch normalization layers, and activation functions. As input, it requires 372-element vector embeddings and processes it through three linear layers with 372, 300, and 200 units, respectively. Each linear layer is followed by batch normalization and then a \emph{Tanh} activation function. Additionally, we also include L1 and L2 regularization across the entire network. We empirically determined that a three-layer network architecture was optimal after testing models (on the validation set) with one to four layers and neuron counts ranging from 100 to 400. Models with one and two layers did not achieve high accuracy, while four-layer models showed similar performance to the three-layer ones without significant improvements.

\subsection{Loss function}
\label{scn: loss_function}

We adopt the pairwise deep supervised hashing (DSH) loss function used in the paper by \citet{liu2016deep}. It is preferred over other loss functions because it enhances the separability of the image feature representation space by incorporating supervised information from input image descriptors. Therefore, image descriptor pairs that are similar are clustered together while dissimilar ones are separated. Additionally, the loss function is designed to regularize the continuous-valued outputs to approximate the desired binary values. The loss function simultaneously learns both the model prediction and the quantization loss in the training process. 

Suppose $\Omega$ represents the COSFIRE descriptor embedding space of the radio galaxy images. Then, the objective is to learn a hash function $F : \Omega \rightarrow \{-1, +1\}^{k}$, which projects the feature embedding space $\Omega$, to a $k$-bit compact binary code space, where $k$ represents the desired dimension of the Hamming space of the binary representations. The mapping must be learned in such a way that visually similar radio galaxy images are assigned binary representations that are close to each other (i.e. small Hamming distance), while dissimilar images are mapped to binary representations that are far apart in the binary code space (i.e. large Hamming distance).

With $N$ training pairs of image feature embedding descriptors and their associated labels ${(D_{i,1}, D_{i,2}, y_i)~|~i = 1, ..., N}: D_1,D_2 \in \Omega$, the respective outputs from the binary network are represented by $b_{i,1}$ and $b_{i,2}$, which belong to $\{-1, +1\}^k$.  The indicator variable $y$ is defined as: $y=0$ if the image-descriptor pairs are similar, and $y = 1 $ if they are dissimilar.

The overall pairwise loss function by \cite{liu2016deep}, is defined as follows: 

\begin{equation}
\begin{aligned}
\mathcal{L}_{DSH} & =\sum_{i=1}^N\left\{\frac{1}{2}\left(1-y_i\right)\left\|\mathbf{b}_{i, 1}-\mathbf{b}_{i, 2}\right\|_2^2\right. \\
& +\frac{1}{2} y_i \max \left(m-\left\|\mathbf{b}_{i, 1}-\mathbf{b}_{i, 2}\right\|_2^2, 0\right) \\
& \left.+ \alpha \left(\left\|\left|\mathbf{b}_{i, 1}\right|-\mathbf{1}\right\|_1+\left\|\left|\mathbf{b}_{i, 2}\right|-\mathbf{1}\right\|_1\right)\right\}
\end{aligned}
\label{eq:loss_function}
\end{equation}

\noindent where $\mathbf{b}_{i, j} \in\{+1,-1\}^k, i \in\{1, \ldots, N\}, j \in\{1,2\}$, while~$\left\|.\right\|_p$,  and $\left|.\right|$ denote the pairwise $\ell_p$-norm, and the elementwise absolute value operation, respectively. The margin threshold parameter is denoted by $m$ ($m > 0$), and finally  $\alpha$ is a weighting parameter that controls the regularization strength.

In equation \ref{eq:loss_function}, the first term encourages similar pairs to have small distances - punishes similar images mapped to different binary representations. The second term encourages dissimilar pairs to have distances greater than the margin $m$, which punishes dissimilar images mapped to a too similar binary representation and only becomes relevant when their Hamming distance falls below the margin threshold $m$. Only those dissimilar pairs having their distance within a radius (margin) are eligible to contribute to the loss function. The third term is the quantization regularization loss: it reduces the discrepancy between the Euclidean and the Hamming code representations by employing a relaxation function \emph{Tanh} which facilitates the thresholding procedure. Moreover, a regularizer is applied to help force the real-valued network outputs from \emph{Tanh} to approach the desired discrete binary-like values (-1,+1).

\subsection{Binarization and image retrieval}
\label{scn: binarization_and_image_trieval}

A trained model is used to generate a real-valued vector for each image in the image retrieval database, including both training and test images. These vectors are then binarized through a thresholding process, where a fixed threshold is applied: elements above the threshold are set to 1, and all others are set to 0. The resulting binary vectors provide compact and efficient representations of the original images. The same encoding process is applied to any given query image, producing its binary representation. These binary vectors are advantageous for image retrieval, as they are more compact and require less storage space compared to the original images or their real-valued feature vectors.

\begin{figure*}
 \centering
 \footnotesize
\includegraphics[width=\textwidth]{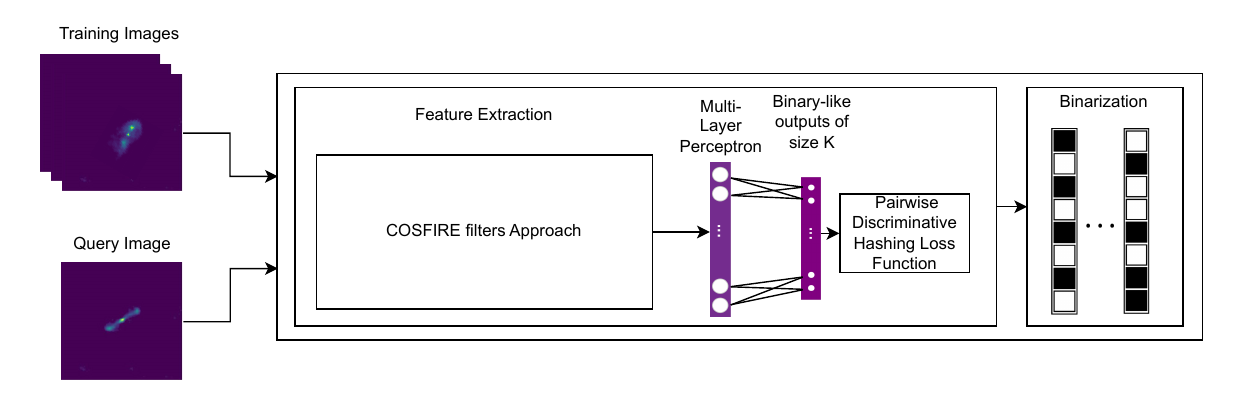}
\caption{A graphic representation of the proposed COSFIRE-based pipeline for radio galaxy image retrieval. It illustrates the feature extraction from the images, and the hashing function process with a pairwise discriminative loss function to obtain binary-like outputs. Finally, it shows the binarization process of the images for the downstream image retrieval task.}
\label{fig: methodology_flow}
\end{figure*}

Moreover, the binary representations enable faster similarity computations, as the Hamming distance between binary vectors can be efficiently calculated using bitwise operations on the hardware. The similarity between binary representations is inversely proportional to the Hamming distance, making it an effective measure for image retrieval tasks, Fig.~\ref{fig: methodology_flow}.

\section{Performance metrics for evaluation}
\label{sec: evaluation}

The Mean Average Precision ($mAP$) is used to evaluate the performance of our proposed methodology \citep{liu2016deep,10265687}.  

\begin{equation}
AP@k = \frac{1}{\text{min(R,k)} }\sum_{i=1}^{k}\text{Precision}(i)\times\text{Rel}(i),
\end{equation}

\begin{equation}
mAP@k = \frac{1}{N_q }\sum_{j=1}^{N_q}{AP@k}_j,
\end{equation}

In this equation, $AP@k$ denotes the average precision at rank $k$ for a single query, where $k$ represents the cutoff rank of the number of top retrieved images under consideration, and $\text{R}$ represents the total number of relevant images for the query. The precision of the top $i$ ranked reference images is given by $\text{Precision}(i)$, while $\text{Rel}(i)$ is a binary indicator variable that takes the value of 1 if the $i$th image is relevant (same label as the query image) and 0 otherwise. The mean average precision ($mAP@k$) is subsequently derived by computing the mean of all $AP@k$ values obtained across the entire set of $N_q$ query images.

The $mAP$ serves as a benchmark for assessing the efficacy of image retrieval systems. It considers both the relevance and ranking of the retrieved images in relation to a set of ground-truth relevant images for each query. A higher $mAP$ score indicates that the system effectively retrieves a substantial proportion of relevant images and ranks them highly within the retrieved set.

\section{Experiments and results}
\label{sec: experiments_and_results}

This section presents detailed analyses of our investigation into radio galaxy image retrieval based on our experimental findings.  Furthermore, we provide a comparison between our proposed method and an existing approach in the field of radio galaxy image retrieval. This comparison encompasses both the effectiveness of image retrieval and the computational complexity of each methodology on the same data set.

\subsection{Performance}

\begin{figure*}
    \centering
    \footnotesize
    \begin{tabular}{cc}
        \includegraphics[trim={15.0 0.7cm 15.0 0.5cm},clip,width=0.48\textwidth]{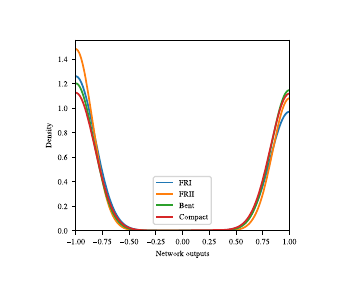} &
        \includegraphics[trim={0 0.5cm 0 0.5cm},clip,width=0.48\textwidth]{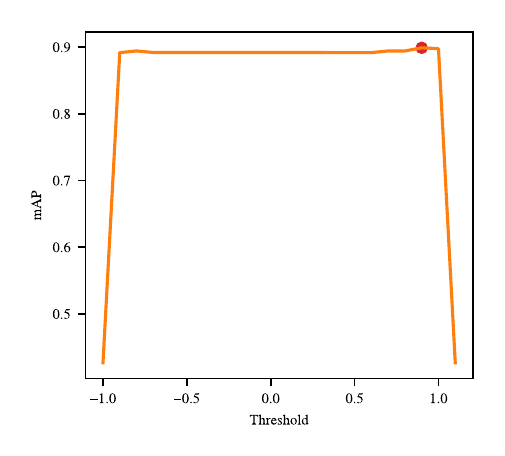}\\     
    \end{tabular}
   
    \caption{ (Left) The Kernel Density Estimation  curves for each galaxy class (FRI, FRII, Bent, and Compact) exhibit bimodal distributions, with peaks around the extreme values of -1 and 1. The model prediction outputs are binary-like, which is ideal for the downstream image retrieval task as it minimizes the quantization error. (Right) The line curve shows the $mAP$ as a function of the threshold values ranging from -1 to 1 applied during the binarization process phase. The dot marker indicates the threshold of 0.9, yielding the maximum $mAP$ of 89.88\% on the validation data.}
    \label{img:Density_plot_and_thresholding_plot}
\end{figure*}

The binarization process was performed by quantizing the network outputs. Fig. \ref{img:Density_plot_and_thresholding_plot}(a) shows their distribution, using fixed values ranging from -1 to 1, with a step size of 0.1. We then evaluated the performance of the image retrieval pipeline using the $mAP$ metric described in Section \ref{sec: evaluation}. The test set served as the query database, while the training set was used as the reference database. Using the validation data set, we identified the threshold that yielded the highest $mAP$, which was subsequently applied to the test data, see Fig. ~\ref{img:Density_plot_and_thresholding_plot}(b). This threshold value also functions as a hyperparameter in our workflow among other hyperparameters.  

The first step involved determining the right configuration of the COSFIRE filters. Based on the findings of \citep{10.1093/mnras/stae821}, there are 26 different sets of COSFIRE configurations, none of which show statistically significant differences in performance. Consequently, we evaluate our approach using all 26 COSFIRE configurations. Utilizing the validation data sets and a MLP hashing architecture, we embarked on a series of experiments to determine the optimal model configuration. These experiments entailed a comprehensive exploration of various hyperparameters pertaining to both the MLP architecture and the pairwise loss function. The specifics of these hyperparameters are shown in Table \ref{tab:hyperparameters}. For each set of validation descriptors, the hyperparameter grid search space resulted in 1,296 distinct experiments. Given that 26 separate experiments were conducted for each hyperparameter configuration, we generate a matrix of $1296\times26$ $mAP$ values, corresponding to each of the 26 COSFIRE configurations and every unique hyperparameter combination.

\begin{table}
\footnotesize
\centering
\caption{Hyperparameter search space for MLP hashing network. This table presents the set of hyperparameters considered to determine the optimal configuration of the MLP hashing architecture. The bottom two rows indicate parameters used in the loss function. The rest are for the MLP architecture. }
\label{tab:hyperparameters}
\begin{tabular}{@{}p{3.0cm}p{5cm}@{}}
\toprule
\textbf{Parameter name} & \textbf{Values} \\
\midrule 
Bit size & $\{16, 24, 32, 40, 48, 56, 64, 72, 80\}$ \\
\addlinespace
Learning rate & $\{0.1, 0.01\}$ \\
\addlinespace
Batch size & $\{32, 48, 64\}$\\ 
\addlinespace
L1 regularization & \{0, 1e-8\}\\
\addlinespace
L2 regularization &\{0, 1e-8\}\\
\addlinespace
\midrule
Margin $m$ & $\{24, 36, 48\}$\\
\addlinespace
Regularization strength $\alpha$ & \{1e-3, 1e-5\}\\

\bottomrule
\end{tabular}
\end{table}

Prior to the main analysis, we divided the matrix according to the bit size hyperparameter, as our analysis is designed to investigate whether there is enhancement in the binary representations and hence the performance as we increase the bit size. This examination aims to evaluate the possible computational benefits that a lower bit-size may yield. Considering the bit sizes of 16, 24, 32, 40, 48, 56, 64, 72, and 80, we extract the corresponding eight submatrices of size $144 \times 26$. The first submatrix corresponds to the bit size 16, the second to 24, and so forth for the remaining bit sizes.

After analyzing the results for each bit size, we computed the average score for each hyperparameter set (represented by the 144 rows in the matrix). To identify the optimal configurations, we applied the right-tailed Mann-Whitney U test (also known as the Wilcoxon rank sum test) \citep{wilcoxon1945some,mann1947test,mcknight2010mann} to compare the row with the highest average score to all other rows. The Mann-Whitney test was chosen over the standard $t$-test because the row vectors had a length of 26 (n = 26), which according to the Central Limit Theorem, is insufficient to assume a normal distribution as required by the $t$-test \citep{moore1989introduction}. The test revealed that for the 16-bit sub matrix, for instance, 14 rows did not exhibit statistically significant differences from the row with the highest average. This process was repeated for each sub-matrix corresponding to different bit sizes.

\begin{figure}
 \centering
 \footnotesize
\includegraphics[trim={0.45cm 0.45cm 0.45cm 0.45cm},clip,width=\columnwidth]{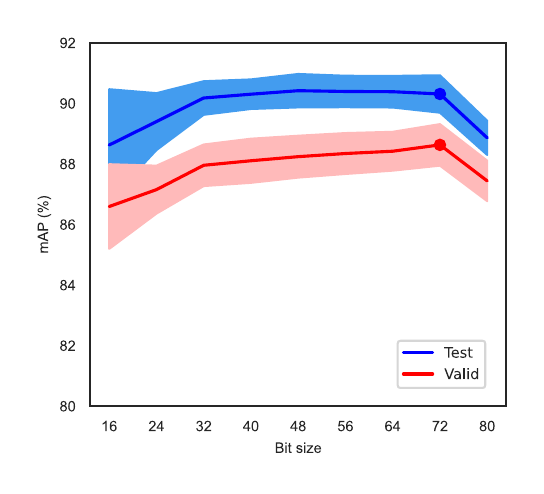}
\caption{The $mAP$ achieved on the validation and test sets as a function of the bit size (16, 24, 32, 40, 48, 56, 64, 72, and 80). The line plots display the average $mAP$ obtained from experiments utilizing the 26 COSFIRE configurations (refer, Section ~\ref{sec:COSFIRE_descriptors_selection}), with a filled envelope representing the standard deviations. The two dot markers indicate the maximum $mAP@100$ achieved with a 72-bit size in reference to the validation data (Table \ref{tab:original_dataset_distrb}).}
\label{img:bit_performance}
\end{figure}

A comprehensive analysis of various bit sizes revealed that a 72-bit hash size was the optimal configuration for our application, yielding the highest $mAP@100$ of 88.63 ± 0.69\% on the validation set. The corresponding test accuracy at 72 bits was 90.31 ± 0.62\% ($mAP@100$). Table \ref{tab:validation_results} and Fig.~\ref{img:bit_performance} present the validation and test results for each bit size. It is noteworthy that while the 72-bit configuration delivered the highest accuracy in the validation set, the improvement over smaller bit sizes - specifically those above 40 bits - is minimal. For example, a 40-bit hash size achieved a validation accuracy of 88.11 ± 0.73\% ($mAP@100$). Therefore, if the goal is to balance accuracy with computational efficiency, a lower bit size such as 40 bits, may offer a more suitable trade-off.

\begin{table}
\footnotesize
\centering
\caption{\textbf{Comparison of bit size with their corresponding validation and test $mAP@100$.}}
\label{tab:validation_results}
\begin{tabular}{@{}p{2.5cm}p{2.5cm}p{2.5cm}@{}}
\hline
\textbf{Bit size} & \textbf{$mAP$ validation (\%)} & \textbf{$mAP$ test (\%)} \\
\hline
16	&   86.59	& 	88.62 \\
\addlinespace
24	& 	87.15	& 	89.40 \\
\addlinespace
32	& 	87.96	& 	90.18 \\
\addlinespace
40	& 	88.11	& 	90.30 \\
\addlinespace
48	& 	88.24	& 	90.42 \\
\addlinespace
56	& 	88.35	& 	90.39 \\
\addlinespace
64	& 	88.42	& 	90.39 \\
\addlinespace
72	& 	88.63	& 	90.31 \\

\addlinespace
80	& 	87.44	& 	88.86 \\

\hline
\end{tabular}

\end{table}

\subsection{Comparison with previous work}

The comparative methodology employs the DenseNet-161 architecture, as implemented in the study by \citet{10265687}. The paper was selected for comparison for two reasons: it provides state-of-the-art results on the data used in the image retrieval task at hand and offers a deep hashing end-to-end approach for image retrieval. It serves as a suitable reference for comparison, given that the data set utilized in the paper is identical to the one used in this project. In contrast, our paper presents an alternative feature-based pipeline. Notably, in applying the deep hashing approach, we made minor modifications—specifically adjusting the loss function and the hashing layer—to enable a more precise comparison. The new loss function is the same one used in our COSFIRE-based approach (equation \ref{eq:loss_function}). This enhancement was essential to ensure a fair comparison between the two methodologies. \citet{10265687} utilized a DenseNet-161 model, which was pretrained on the ImageNet data set, subsequently fine-tuned using the astronomical data set outlined in Table \ref{tab:original_dataset_distrb}. The final layer of the architecture is a fully connected layer. This information helps the layer to perform the hashing process effectively. Contrary to the sigmoid function mentioned in the original paper, the \emph{Tanh} activation function is applied in the hash layer, ensuring all output values are restricted within the interval [-1,1].

\subsubsection{Model robustness}

Our experimental results demonstrate that the COSFIRE approach outperformed DenseNet-161, achieving an average $mAP@(k=R)$ of 90.51\%, using a bit size of 72 as determined from the validation set, compared to 89.82\% for DenseNet-161 (bit size of 72). The performance breakdown for each radio galaxy category is presented in Table ~\ref{tab:mAP_at_R_equals_K}. The $mAP@(k=R)$ represents a specialized variant of the mean average precision calculated when the cutoff rank $k$ is set equal to $R$, the total number of relevant images. This configuration ensures a balanced consideration of precision and recall, as the number of retrieved images precisely matches the number of relevant images. This alignment provides a robust and accurate evaluation of the system's performance.

\begin{table}
\footnotesize
\centering
\caption{\textbf{The mean average precision (\%) per class when R = K for both COSFIRE and DenseNet-161 approaches on the test data.}}
\label{tab:mAP_at_R_equals_K}
\begin{tabular}{@{}p{1.5cm}p{2.4cm}p{1.6cm}p{1.7cm}@{}}
\hline
\textbf{Galaxy type} & \textbf{Relevant images (R)} & \textbf{COSFIRE} & \textbf{DenseNet-161} \\
\hline
Bent	&   305	&  90.01 & 	89.79 \\
\addlinespace
FRII	& 	434	&  90.69 & 	90.37 \\
\addlinespace
FRI  	& 	215	&  90.67 & 	89.56 \\
\addlinespace
Compact	& 	226	&  90.68 & 	89.57 \\
\hline
\textbf{Average}  &   & \textbf{90.51}    & 89.82  \\
\hline
\end{tabular}

\end{table}

\begin{figure}
 \centering
 \footnotesize
\includegraphics[trim={0.5cm 0.45cm 0.5cm 0.55cm},clip,width=\columnwidth]{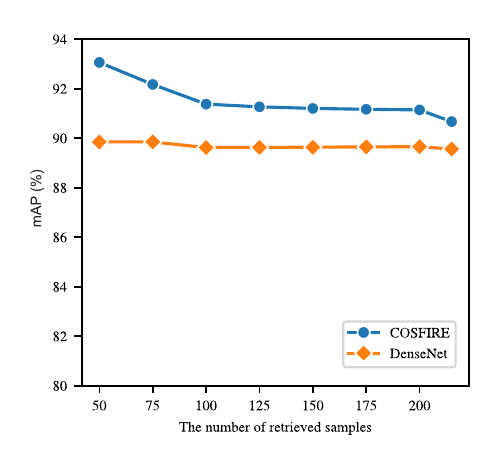}
\caption{ $mAP$ against the number of retrieved samples for COSFIRE and DenseNet-161 models for the top 215 (size of smallest radio galaxy class). Both models show robust performance with the COSFIRE approach having a better retrieval $mAP$ score.}
\label{fig:map_precision_recall_number_images}
\end{figure}

We further examine how the image retrieval performance of the models compares as the number of top $N$ images increases. The $N$ is capped at the point where $K=R$ for the smallest radio galaxy class\footnote{In this particular case, $N=215$, which is the size of the FRI class in the training data}, to ensure that each query is evaluated based on retrieving as many images as there are relevant ones in the database. As depicted in Fig.~\ref{fig:map_precision_recall_number_images}, the COSFIRE model consistently outperforms DenseNet-161 across all sample sizes. The $mAP$ of COSFIRE starts at approximately 93\% for 50 samples and gradually decreases to 90.5\% for 200 samples. In contrast, DenseNet-161 maintains a relatively stable $mAP$ of around 89.5-90\%, demonstrating minimal variation with changes in sample size.

\subsubsection{Computational complexity}

Computational efficiency is a critical factor in model selection. Evaluating the computational complexity of the COSFIRE and DenseNet-161 approaches involves utilizing the Floating Point Operations (FLOPs) metric, which measures processor requirements by counting the number of floating-point operations needed to complete a task. We adopt and extend the detailed calculations from \citet{10.1093/mnras/stae821}, focusing on the inference phase. According to this study, DenseNet-161 requires approximately 15.6 GFLOPs, while the COSFIRE approach needs around 1.1 GFLOPs. The COSFIRE FLOPs are calculated by summing the operations required for deriving COSFIRE descriptors and those needed in the MLP architecture (Table ~\ref{tab:flops}): 1,139,692,788 + 374,572 = 1,140,067,360 FLOPs (1.1 GFLOPs). These calculations reveal that the COSFIRE approach demonstrates superior computational efficiency compared to DenseNet-161. The efficiency of the COSFIRE approach is attributed to its use of separable filters, where 2D Gaussian functions are split into two 1D filters, significantly reducing the number of operations. Furthermore, COSFIRE employs optimization strategies such as eliminating redundant computations, sharing configurations among filters, and utilizing pre-computed response maps stored in hash tables \citep{10.1093/mnras/stae821}. This approach not only reduces computational costs but also maintains high performance while offering flexibility in filter customization based on specific needs.

 \begin{table}
 \caption{\textbf{FLOPs calculation for each component in the MLP Architecture. Note that $m$ and $n$ represent the data matrix input size and output size respectively.}}
 \footnotesize
\centering
\begin{tabular}{@{}p{3.8cm}p{0.9cm}p{3cm}@{}}
\hline
\textbf{Component} & \textbf{Formula} & \textbf{FLOPs} \\
\hline
\textbf{Linear Layer 1:} m = 372, n = 300 & $2mn$ & $2\times372\times300 = 223,200$ \\
\hline
\textbf{Batch Normalization 1:} n = 300 & $4n$ & $4 \times 300 = 1,200$ \\
\hline
\textbf{\emph{Tanh} Activation 1:} n = 300 & $n$ & $300$ \\
\hline
\textbf{Linear Layer 2}: m = 300, n = 200 & $2mn$ & $2 \times 300 \times 200 = 120,000$ \\
\hline
\textbf{Batch Normalization 2:} n = 200 & $4n$ & $4 \times 200 = 800$ \\
\hline
\textbf{\emph{Tanh} Activation 2:} n = 200 & $n$ & $200$ \\
\hline
\textbf{Linear Layer 3:} m = 200, n = 72 & $2mn$ & $2 \times 200 \times 72 = 28,800$ \\
\hline
\textbf{\emph{Tanh} Activation 3:} n = 72 & $n$ & $72$ \\
\hline
\textbf{Total number of FLOPs} & & 374, 572\\
\hline
\end{tabular}

\label{tab:flops}
\end{table}

\subsubsection{Class Separability Analysis}

\begin{figure*}
    \centering
    \footnotesize
    \includegraphics[trim=1cm 0cm 0.1cm 7.5cm,clip,scale=0.95]{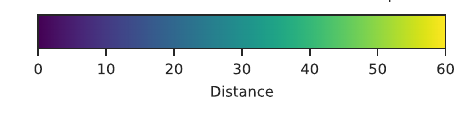} \\

    \begin{tabular}{cc}
        \includegraphics[trim=0.1cm 2cm 0.1cm 0.2cm,clip,width=0.48\textwidth]{hamming_distance_matrix_heatmap_cosfire_test_data.pdf} &
        \includegraphics[trim=0.1cm 2cm 0.1cm 0.2cm,clip,width=0.48\textwidth]{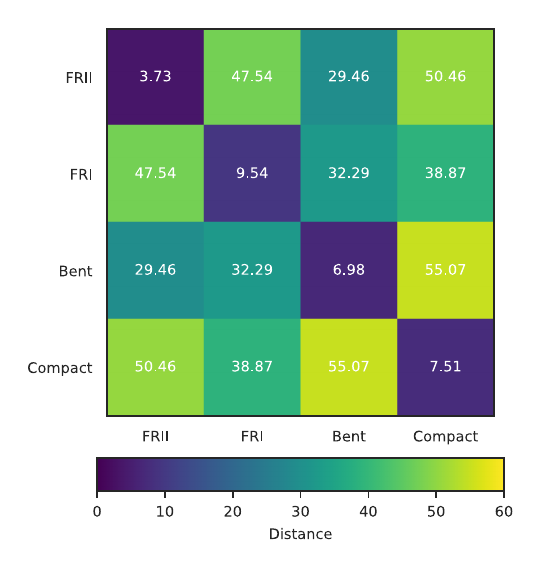} \\
        ~~~~~~~~~~~~~(a) COSFIRE (test data) & ~~~~~~~~~~~~~~~~(b) DenseNet-161 (test data) \\[6pt]
        
        \includegraphics[trim=0.1cm 2cm 0.1cm 0.2cm,clip,width=0.48\textwidth]{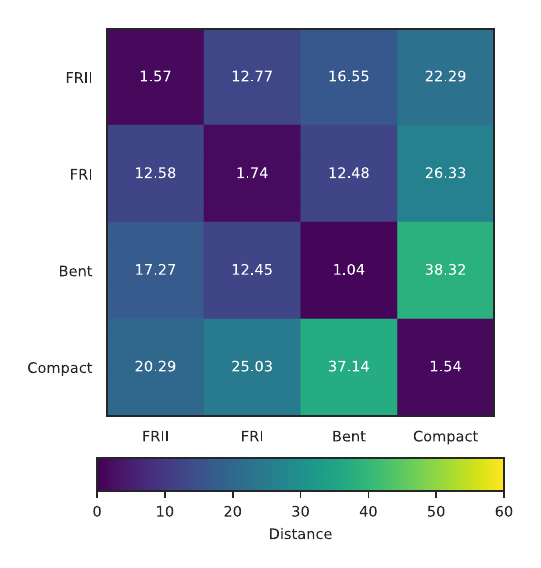} &
        \includegraphics[trim=0.1cm 2cm 0.1cm 0.2cm,clip,width=0.48\textwidth]{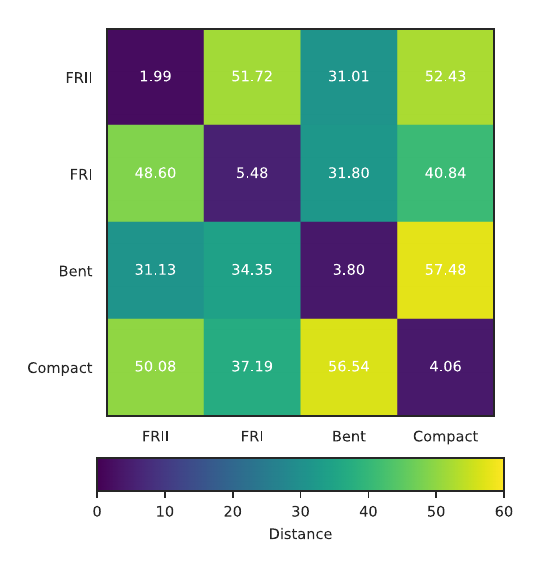} \\
        ~~~~~~~~~~~~~(c) COSFIRE (test vs. training data) & ~~~~~~~~~~~~~(d) DenseNet-161 (test vs. training data) \\
    \end{tabular}
    
    \caption{Intra- and inter-class Hamming distances for the COSFIRE and DenseNet-161 approaches across the four radio galaxy classes in our dataset. Matrices (a) and (b) display distances within the test dataset for COSFIRE and DenseNet-161, respectively, while matrices (c) and (d), illustrate distances between the test and training data, where Hamming distances are computed between each test image and all training images. The diagonal elements represent intra-class distances, and off-diagonal elements show inter-class distances. Notably, the COSFIRE approach shows a clear reduction in intra-class distances and their variance across the four galaxy types, particularly evident in (a) and (c). Such compact class representation contributes to enhanced mAP performance when using test images as queries and training images as the reference database.}
    \label{fig:intraclass_distances}
    
\end{figure*}

To understand the source of COSFIRE's improvement, we evaluated class separability in our dataset by analyzing intra-class and inter-class Hamming distances using the COSFIRE and DenseNet-161 models. Intra-class distances represent the average distance between all instances of the same class, whereas inter-class distances measure the average distance between all instances of different classes. Fig.~\ref{fig:intraclass_distances} illustrates these distances through matrices, where the leading diagonal represents intra-class distances and the off-diagonal elements show inter-class distances. We examined two scenarios: distances within the test set (Fig.~\ref{fig:intraclass_distances}a and b) and distances between the test and training sets (Fig.~\ref{fig:intraclass_distances}c and d). In the latter case, Hamming distances were computed between each test image and all training images. To quantify separability, we calculated the ratio of mean intra-class to mean inter-class distances. For the test set, COSFIRE yielded a ratio of 0.1088, compared to 0.1641 for DenseNet-161. When comparing test to training data, the ratios were 0.0687 and 0.0867 for COSFIRE and DenseNet-161, respectively. These lower ratios indicate greater class separability, suggesting that COSFIRE effectively achieves a greater separation between classes in our dataset. In particular, COSFIRE produces significantly more compact class representations than DenseNet-161.

\section{Discussion}
\label{sec: discussion}

The COSFIRE approach has proven to be a highly effective technique for generating robust and discriminative image descriptors for both classification and retrieval tasks in astronomical contexts. Its trainability allows adaptation to specific shapes and orientations of prototype patterns, while its inherent explainability makes it ideal for capturing intricate morphological details in celestial sources. COSFIRE's tolerance to rotation, achieved by incorporating responses from multiple rotated versions of training images, is particularly valuable for handling celestial objects in various orientations due to observational perspectives. These properties are crucial when dealing with astronomical data \citep{1998ASPC..145..429P,scaife2021fanaroff,brand2023feature}.

In our comparative analysis, COSFIRE consistently outperforms DenseNet-161, especially within the top 100 retrieved images, by retrieving fewer irrelevant images in the early ranks. This capability is critical for maintaining high precision in astronomical applications. While DenseNet-161 tends to introduce irrelevant images earlier in the ranking, COSFIRE maintains higher accuracy. Additionally, the COSFIRE methodology demonstrated robustness in maintaining a low intra-class to inter-class distance ratio. It effectively created significant separation between radio galaxy source classes despite their intricate and complex morphological characteristics, while simultaneously keeping instances within the same class closely grouped. This performance suggests that COSFIRE can distinguish between different classes of radio galaxies while preserving the similarity of objects within each class, even when faced with the challenging variability inherent in radio galaxy morphology.

Deep learning methods for CBIR have significantly advanced the field by enabling the automatic extraction of high-level features from images. However, these methods often require substantial computational resources, including high memory and processing power \citep{sayed2023deep,10265687,ghozzi2024deep}. The computational intensity of these models arises from their architecture, which includes numerous layers and parameters that facilitate the learning of complex features. The computational demands of deep learning models, such as DenseNet-161, present a challenge, particularly in resource-constrained environments. As observed in this work, DenseNet-161 requires approximately 15.6 GFLOPs, making it $\sim$14 times more resource-intensive compared to the COSFIRE approach, which achieves comparable results with only 1.1 GFLOPs. This remarkable reduction in computational complexity makes COSFIRE a more viable option for scenarios where efficiency and low resource usage are crucial.

Representations are essential for the success of machine learning and deep learning algorithms in astronomy, a data-rich field. As demonstrated in this work and by \citet{walmsley2022practical}, learning meaningful feature representations enables the development of models capable of performing a broad spectrum of downstream astronomical tasks, including classification, image retrieval, and anomaly detection. To the best of our knowledge, this study and our prior work \citep{10265687} are pioneering efforts in applying machine and deep learning techniques to data within the field of radio astronomy.

\section{Conclusions}
\label{sec: conclusion}
This paper presents COSFIRE filters as an effective method for generating image descriptors that excel in MLP-based image hashing. The COSFIRE-MLP approach achieved an $mAP@(k=R$) of 90.51\%, outperforming a DenseNet-161 transfer learning approach that yielded an $mAP@(k=R$) of 89.82\% on the same data set. Additionally, COSFIRE required 14 times fewer FLOPS, making it significantly more computationally efficient. The $mAP@(k=R)$ is the mean average precision calculated when $k$, the cutoff rank, is set equal to $R$, the total number of relevant items. Our model effectively learns discriminative features, enabling efficient retrieval of morphologically similar radio sources.

Compared to deep learning methods, COSFIRE offers advantages in efficiency, scalability, robustness, and high-quality image retrieval. This approach is well-suited for large-scale systems, allowing astronomers to quickly find similar celestial objects in extensive archives. By enhancing data archive accessibility, COSFIRE also provides an intuitive and accessible method for disseminating scientific knowledge within the astronomy community.

This study advances the field of astronomy by introducing a novel approach to retrieving images of celestial sources. As next-generation high-resolution telescopes like MeerKAT, LOFAR, and SKA generate increasingly detailed images of the sky, our future research will investigate the applicability of our method to these new data sets and explore the potential for cross-survey predictions.

\section*{Acknowledgements}

\noindent Part of this work is supported by the Foundation for Dutch Scientific Research Institutes. This work is based on the research supported in part by the National Research Foundation of South Africa (grant numbers 119488 and CSRP2204224243).

\noindent The financial assistance of the South African Radio Astronomy Observatory (SARAO) towards this research is hereby acknowledged (www.ska.ac.za)

\noindent We thank the Center for Information Technology of the University of Groningen for their support and for providing access to the Hábrók high performance computing cluster.



\section*{Data Availability}

The data set is available at: \url{https://github.com/kiryteo/RG\_Classification\_code}.



\bibliographystyle{mnras}
\bibliography{bibliography} 







\label{lastpage}
\end{document}